\documentclass[a4paper,10pt,superscriptaddress,showpacs,showkeys,nofootinbib,aps,reprint]{revtex4-1}
\usepackage{amsmath,amssymb}
\usepackage[final]{graphicx}
\usepackage[english]{babel}
\usepackage[utf8]{inputenc}
\usepackage{hyperref}% add hypertext capabilities
\usepackage{epstopdf}

\graphicspath{{./}}
\makeatletter
\renewcommand{\p@subsection}{}
\renewcommand{\p@subsubsection}{}
\makeatother

    \def\<{\langle}
    \def\>{\rangle}

    \def\vphi{\varphi}
    \def\pt{\partial}

  \newcommand{\eq}[1]{
    \begin{equation}
    {#1}
    \end{equation}}

    \newcommand{\nmq}[1]{
    \begin{multline}
    #1
    \end{multline}}

\begin{document}
\title{Tunneling current noise in the fractional quantum Hall effect: when the effective charge is not what it appears to be}
\author{Kyrylo Snizhko}
\affiliation{Faculty of Physics, Taras Shevchenko National University of Kyiv, Kyiv, 03022, Ukraine}
\affiliation{Shortly after leaving Department of Physics, Lancaster University, Lancaster, LA1~4YB, UK}

\begin{abstract}
Fractional quantum Hall quasiparticles are famous for having fractional electric charge. Recent experiments report that the quasiparticles' effective electric charge determined through tunneling current noise measurements can depend on the system parameters such as temperature or bias voltage. Several works proposed to understand this as a signature for edge theory properties changing with energy scale. I consider two of such experiments and show that in one of them the apparent dependence of the electric charge on a system parameter is likely to be an artefact of experimental data analysis. Conversely, in the second experiment the dependence cannot be explained in such a way.
\end{abstract}


\maketitle

\section{INTRODUCTION}

The fractional quantum Hall effect (FQHE, fractional QHE) is a field of intensive research nowadays, with one of the main reasons for that is its supporting of quasiparticle excitations with unusual properties. Quasiparticles in the FQHE have the electric charge which is a fraction of the electron charge, and are predicted to have other unusual properties such as anyonic or even non-Abelian statistics. The quasiparticles obeying the non-Abelian statistics would potentially allow for performing topologically protected quantum computations (TPQC) (i.e., quantum computations in which qubits are protected from decoherence by "topological order" of the system) \cite{TopolQuantCompFQHE}. Therefore, finding the properties of quasiparticles in different FQHE states is an important task.

Measuring tunneling current noise is a powerful method for finding the properties of quasiparticle excitations in the FQHE, in particular the tunneling quasiparticle electric charge. In the regime of weak tunneling of quasiparticles the tunneling current shot noise is proportional to the tunneling current itself, the proportionality coefficient, called the Fano factor, is the tunneling quasiparticle charge \cite{Kane1994}. If several quasiparticles contribute to the tunneling processes, then the Fano factor is some average of the quasiparticles' charges. It is in this way that the first confirmation was given for the fractional charge of quasiparticles in the FQHE with filling factor $\nu = 1/3$ \cite{FracChargeObs_Heiblum, FracChargeObs_Glattli}.

Some of more recent experiments \cite{Griffiths2000,Chung2003,Dolev2008,HeiblumForFerraro,Dolev2010,HeiblumExp} that studied more complicated FQHE states report that the "effective charge" determined from tunneling current noise depends on external parameters such as temperature \cite{Chung2003, HeiblumForFerraro}, bias voltage across the tunneling contact \cite{Dolev2010}, other system parameters \cite{HeiblumExp}. Three distinct mechanisms proposed recently can contribute to evolution of the Fano factor. Two of them assume that the FQHE edges behave differently at different energy scales: either due to energy cutoffs of edge transport channels \cite{Ferraro2008, Ferraro, Carrega2011} or due to edge reconstruction~\cite{MeirGefen23EdgeReconstruction}. The third one, considered in Ref.~\cite{ShtaSniChe_2014}, assumes dependence of quasiparticle tunneling amplitudes on experimental parameters, which can change relative importance of different quasiparticles' contributions.

However, there is a subtlety regarding the data analysis in experimental works. Namely, experimentalists \cite{FracChargeObs_Heiblum, FracChargeObs_Glattli, Griffiths2000,Chung2003,TunnellingRate2,Dolev2008,HeiblumForFerraro,Dolev2010,HeiblumExp} tend to use for analysis a formula that is not based on a realistic FQHE model but is a generalization of a formula which can be derived for free electrons. As it was analyzed in Ref.~\cite{SaleurGlattliCothangentNoiseValidity} in the case of $\nu = 1/3$, the formula used by experimentalists can agree well with the exact theory under certain conditions, but deviates from the exact theory otherwise. This can give rise to misinterpretations of experimental data. In particular, the "effective charge" obtained this way is not necessarily the same as the Fano factor.

In this work I analyze the results of Ref.~\cite{HeiblumExp} regarding $\nu = 2/3$ and of Ref.~\cite{Chung2003} regarding $\nu = 2/5$. I show that in the former case the data can be explained within the minimal $\nu = 2/3$ FQHE edge model \cite{KaneFisherPolchinski, KaneFisher} without additional structure such as energy cutoffs or edge reconstruction. Therefore, the effective charge dependence on external parameters appears to be a data analysis artefact in this case. In the case of the data of Ref.~\cite{Chung2003} regarding $\nu = 2/5$, the charge dependence on the system temperature cannot be explained in this simple way.

The paper structure is as follows. In section~\ref{Sect_TunnExpScheme} I introduce a general scheme of the experiments I discuss. Then in section~\ref{Sect_TunnExp_Model} I describe a theoretical model that can be used to analyse such experiments. This model is not easy to treat, therefore in section~\ref{Sect_TheorDescr} I discuss the three existing approaches to analysing the model and the experiments: the one based on perturbative treatment of tunneling processes (Sec.~\ref{Sect_TheorDescr_Pert}), the one based on exactly solving the model in the cases when this can be done (Sec.~\ref{Sect_TheorDescr_Exact}), and the one typically used by experimentalists~--- the phenomenological approach (Sec.~\ref{Sect_TheorDescr_Phenom}). Finally, sections~\ref{Sect_Nu_23} and \ref{Sect_Nu_25} present original results of analysing the data regarding $\nu = 2/3$ and $\nu = 2/5$ respectively. Some concluding remarks are made in section~\ref{Sect_Discuss_Concl}.

\section{\label{Sect_TunnExpScheme}Tunneling experiments in the FQHE: a typical scheme}

\begin{figure}[tbp]
  \center{\includegraphics[width=1\linewidth]{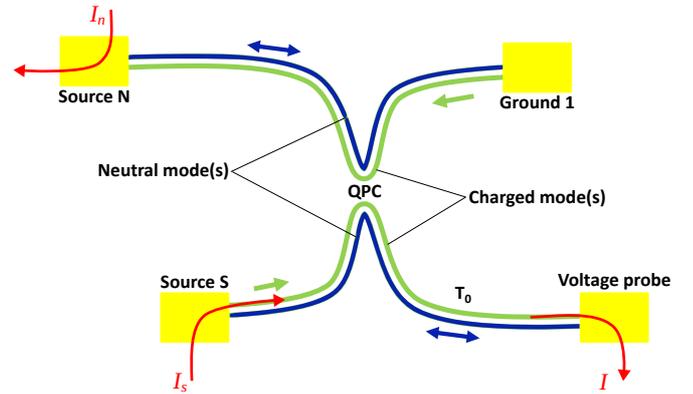}}
  \caption{\textbf{A typical experiment scheme}. (Color online). Two FQHE edges form a quantum point contact (QPC) at which quasiparticles can tunnel between the edges. The Ohmic contact \textit{Ground~1} is grounded. \textit{Source~N} and \textit{Source~S} are used to inject some electric current into the system. Measurement of the electric current and its noise is performed at \textit{Voltage~probe}. $T_0$ is the system and its environment temperature when currents $I_s$, $I_n$ are not injected.}
  \label{fig_exp_scheme}
\end{figure}

The typical scheme of the experiments I am going to discuss is presented in Fig.~\ref{fig_exp_scheme}.

There are two FQHE edges (upper and lower) along which transport of electric charge and of heat can occur, the rest of the sample is insulating. Each edge contains at least one "charged mode"~--- the channel, excitations in which carry electric charge and are responsible for charge transport. The transport channels are chiral, i.e., excitantions in a channel can propagate in one direction only. If there are several charged modes in an edge, I assume that all of them flow in one direction. Apart from the charged modes the edges can support "neutral modes". These are transport channels that do not carry electric charge. They can, however, transport heat, spin etc. The neutral modes can be absent at all, there can be one or several of them. Neutral modes are also chiral. Some of the neutral modes can flow in the same direction as the charged ones, some~--- in the opposite direction.

The upper and lower edges come close together at the quantum point contact (QPC) where tunneling of quasiparticles between the edges can take place. Apart from the QPC, the edges are separated and do not interact with each other.

Experimental equipment is connected to the system through four Ohmic contacts (yellow rectangles). \textit{Ground~1} contact is grounded. \textit{Source~S} is used to inject electric current $I_s$ into the lower edge. \textit{Voltage~probe} is used to measure the current $I$ flowing into it, and its noise. If no tunneling takes place at the QPC, then $I = I_s$ and the noise of $I$ is just the Johnson-Nyquist noise. However, if there is tunneling at the QPC, then both $I$ and its noise carry information about the tunneling processes. Finally, \textit{Source~N} is used to inject current $I_n$ into the system. As one can see from the scheme, the electric current itself does not flow into the system. However, its injection can excite the neutral modes of the upper edge, and if some of them flow opposite to the charged mode, they can influence the tunneling processes at the QPC.

\section{\label{Sect_TunnExp_Model}Tunneling experiments in the FQHE: the model}

In this section I briefly outline the standard model for analyzing the experiments described in the previous section.

The model contains three distinct ingredients: single edge model (to describe each of the two edges), tunneling processes model, and a model for interaction of the Ohmic contacts with the edge.

Here I consider the case of Abelian edge theories. The non-Abelian ones can be considered similarly, but I do not analyze them in this paper. For a general discussion of how the FQHE edge theories are constructed see Ref.~\cite{Frohlich_CosetHall}. A single Abelian edge can be described in terms of $N$ bosonic fields $\vphi_i$ with $i = 1,...,N$, one for each edge mode. The action for the fields is\footnote{In this section I put $e = \hbar = k_B = 1$ unless the opposite is stated explicitly. Here $e$ is the elementary charge, $\hbar$ is the Planck constant, $k_B$ is the Boltzmann constant.}
\eq{\label{action}
S = \frac1{4\pi} \int dx dt \sum_m \Bigl( -\chi_m \partial_x\vphi_m \partial_t \vphi_m - v_{m}(\partial_x\vphi_m)^2 \Bigl),
}
where $\chi_m = \pm1$ determine chiralities of the modes (plus for counterclockwise-movers and minus for clockwise-movers\footnote{Clockwise-movers are left-movers at the lower edge and right-movers at the upper edge. Correspondingly, the counterclockwise-movers are the right-movers at the lower edge and left-movers at the upper edge.}), and $v_{m} > 0$ are the modes' propagation velocities. Without loss of generality, I put $\chi_m = +1$ for $m = 1,..., N_{l}$ and $\chi_m = -1$ for $m = N_{l}+1, ..., N$ ($N_l$ is thus the number of counterclockwise-moving modes).

The electric current $J^\alpha$ ($J^0$ is the electric charge density, $J^1$ is the electric current flowing along the edge) has the form
\eq{\label{el_current operator_2} J^\alpha = \frac{1}{2\pi}\sum_m q_m\varepsilon^{\alpha\beta}\pt_\beta\vphi_m,}
where the symbol $\varepsilon^{\alpha\beta}$ denotes the fully antisymmetric tensor with $\alpha, \beta$ taking values $t$ and $x$ (or 0 and 1, respectively) and $\varepsilon^{t x} = \varepsilon^{0 1} = 1$. The numbers $q_i$ should satisfy the constraint \cite{WenCLL, FrohlichAnomaly}
\eq{\label{anomaly_cancellation}\sum_m \chi_m q_m^2 = \nu,}
where $\nu$ is the filling factor.

As it was mentioned in the previous section, I assume that all the modes that carry electric charge flow in one direction. Formally this means that $q_m = 0$ for $m = N_{l}+1, ..., N$, i.e., only counterclockwise-propagating modes can carry electric charge.

The quantized fields $\vphi_m$ obey the commutation relations
\eq{\label{comm_relations}[\vphi_m(x,t),\vphi_{m'}(x',t')] = -i \pi \mathrm{sgn}(X_m-X_m')\,\delta_{m,m'},}
where $X_m = -\chi_m x+v_m t$.

It is convenient to introduce local quaiparticle operators
\eq{\label{vertex_quasi}V_{\mathbf{g}}(x,t) = \left( \frac{L}{2\pi} \right)^{-\sum_m g_m^2/2} :\exp\left(i\sum_m g_m\vphi_m(x,t)\right):.}
These operators can be not used when describing transport along a single edge, but are important for tunneling processes. Here $L$ is the edge length, $: ... :$ stands for the normal ordering, $\mathbf{g} = (g_1,...,g_N)$, and $g_m\in\mathbb{R}$ are the quasiparticle quantum numbers. The quasiparticles' quantum numbers are quantized, i.e., the set of allowed vectors $\mathbf{g}$ is discrete. The quasiparticle's two most important quantum numbers, the electric charge $Q$ and the scaling dimension $\delta$, are equal to
\begin{eqnarray}
\label{Charge}
Q &=& \sum_m \chi_m q_m g_m,\\
\label{ScalDim}
\delta &=& \frac 12\sum_m g_m^2.
\end{eqnarray}

Having constructed a single edge theory, one can describe tunneling between the two edges at the QPC as hopping of local quasiparticles from one edge to another. The Hamiltonian for such processes is \cite{WenCLL, Kane1994, Saleur, Saleur2}
\eq{\label{tun_ham} H_{T} = \sum_{\mathbf{g}} \eta_{\mathbf{g}} V_{\mathbf{g}}^{(u)\dag}(0,t)V^{(l)}_{\mathbf{g}}(0,t) + \mathrm{h.c.},}
here the superscripts $(u), (l)$ label quantities relating to the upper and the lower edges respectively, $\eta_{\mathbf{g}}$ are the tunneling amplitudes; for simplicity I have put the position of the QPC to the origin of coordinates. In the limit of large edge length ($L \rightarrow \infty$) the dominant contribution to the tunneling processes comes from the quasiparticles with the smallest scaling dimension $\delta$.\footnote{One can see this from $L^{-\delta}$ factor in Eq.~\eqref{vertex_quasi}. This statement is also confirmed by Monte Carlo simulations \cite{HuntingtonCheianov_TunnAmplitudeMonteCarlo}.} In the following I label such quasiparticle types by $i = 1,...,n$, with the quasiparticle electric charges being $Q_i$ (in the units of the elementary charge $e$), their common scaling dimension being $\delta_i = \delta$, and the full set of quantum numbers being $\mathbf{g}^i$.

The final component is a model for interaction between the Ohmic contacts and the edge. For this work I use the following set of assumptions regarding the interaction. I assume that when an edge mode flows into an Ohmic contact all the excitations are absorbed by the latter, and the state of inflowing modes does not influence the state of modes that flow away from the contact. I also assume that an edge mode emitted by an Ohmic contact, when no current is injected into it, is in thermal equilibrium with the contact and its environment. When an electric current is injected through an Ohmic contact, the only change to the state of the charged mode(s) emanating from the contact is the change of their chemical potentials, so that they carry the injected current. The influence of current injection on the neutral modes should be a matter of separate investigation. For this work I assume that the neutral modes that propagate in the same direction as the charged mode(s) are not influenced by current injection at all, while the counterpropagating neutral modes get heated due to this. Therefore, if counterflowing neutral modes are present in the edge, the temperature of the upper edge near the QPC is $T = \lambda(I_n)T_0$, with $\lambda(I_n) \geq \lambda(0) = 1$. Details of this heating for $\nu = 2/3$ were investigated in Ref.~\cite{ShtaSniChe_2014}.

Before reviewing the existing approaches to solving the model outlined above, I define the observables that are measured in the experiments I consider below.

Current $I$ flowing into \textit{Voltage probe} contact (see Fig.~\ref{fig_exp_scheme}) is equal to $J^1$ component of the current $J^\alpha$, defined in Eq.~\eqref{el_current operator_2}, taken at some point to the right of the QPC along the lower edge. I denote the operator of this current as $\hat{I}(t)$. Then, the average current flowing into \textit{Voltage probe} is $I = \langle \hat{I}(t) \rangle$. It is also convenient to introduce operator $\widehat{\delta I}(t) = \hat{I}(t) - I$.

If there is no tunneling at the QPC, then $I$ is equal to the current $I_s$ injected at \textit{Source S}. As soon as there is tunneling, some part of the quasiparticles will not reach \textit{Voltage probe}, with $I_s - I = I_T$ being the tunneling current. Define the following quantities:
\begin{itemize}
\item transmission rate $t = I/I_s$;
\item tunneling (or reflection) rate $r = I_T/I_s = 1 - t$;
\item measured current noise\footnote{One must be cautious when comparing formulas and data for noise from different articles since there are two conventions regarding the definition of the noise spectral density. While some authors (see, e.g., \cite{Lesovik1989}) use the same definition as I do, others (see, e.g., \cite{MartinLandauer, HeiblumExp}) adopt the definition which is twice as large as the one used here.}
 \eq{S(\omega) = \int\limits_{-\infty}^\infty d\tau \exp\bigl(i\omega\tau \bigl) \frac 12 \Bigl\<\Bigl\{\widehat{\delta I}(0),\widehat{\delta I}(\tau)\Bigl\}\Bigl\>,}
\end{itemize}
where $\{\dots\}$ denotes the anti-commutator.

In what follows I only use the zero-frequency noise $S(\omega = 0)$. It is also convenient to talk about the excess noise
\eq{\label{excess_noise_def}
\tilde{S}(\omega = 0) = S(0) - S_{\mathrm{Nyquist}}(0) = S(0) - \frac{\nu}{2\pi} T_{0},
}
where $T_0$ is the system temperature when no currents are injected.

\section{\label{Sect_TheorDescr}Three approaches to theoretical description of tunneling experiments}

\subsection{\label{Sect_TheorDescr_Pert}Perturbative treatment of tunneling}

The model described in the previous section is hard to solve. Exact solutions are available only in exceptional cases. Therefore, the most generally applicable approach is to treat the tunneling Hamiltonian \eqref{tun_ham} as a small perturbation. Then in the lowest non-trivial order of perturbation theory one obtains the following results \cite{ShtaSniChe_2014, Snizhko2015}:\footnote{Here and in the rest of the paper I restore the elementary charge $e$, the Planck constant $\hbar$, and the Boltzmann constant $k_B$, which I had put to $1$ in section~\ref{Sect_TunnExp_Model}.}
\eq{\label{tunn_coeff_expr}r = \frac{4 e (\pi k_B T_0)^{4\delta - 1}}{I_s \hbar^{4 \delta + 1}} \sum_i \kappa_i G_i,}
\eq{\label{noise_expr}\tilde{S}(0) = \frac{4 e^2 (\pi k_B T_0)^{4\delta - 1}}{\hbar^{4 \delta + 1}} \sum_i \kappa_i F_i,}
\eq{
\label{G_i}
G_i = \sin{2\pi\delta} \int\limits_{0}^{\infty}dt \frac {Q_i \lambda^{2 \delta}\,\sin Q_i j_s t}{(\sinh t)^{2\delta}(\sinh \lambda t)^{2\delta}}
}
\eq{
\label{F_i}
F_i = F^{TT}_{i} \cos{2\pi\delta} - \frac{2}{\pi} F^{0T}_i \sin{2\pi\delta},
}
\eq{
\label{F_TT}
F^{TT}_{i} = Q_i^2 \lim_{\varepsilon \rightarrow +0} \left(\frac{\varepsilon^{1-4\delta}}{1-4\delta} + \int\limits_{\varepsilon}^{\infty}dt \frac{\lambda^{2 \delta}\,\cos Q_i j_s t}{(\sinh t)^{2\delta}(\sinh \lambda t)^{2\delta}} \right)
}
\eq{
\label{F_0T}
F^{0T}_{i} = \int\limits_{0}^{\infty}dt \frac {Q_i^2 \lambda^{2 \delta}\, t\cos Q_i j_s t}{(\sinh t)^{2\delta}(\sinh \lambda t)^{2\delta}}
}
\eq{
\label{j_s}
j_s = \frac{I_s}{I_0},\quad I_0 = \nu \frac{e}{h} \pi k_B T_0,
}
where $T_0$ is the equilibrium system temperature, $I_s$ is the current injected into \textit{Source S}, $\lambda = \lambda(I_n)$ is related to the upper edge heating due to injection of current $I_n$ (the upper edge temperature at near the QPC is $T = \lambda T_0$), $e$ is the elementary charge, $h = 2 \pi \hbar$ is the Planck constant, $k_B$ is the Boltzmann constant, $\nu$ is the filling factor, $\kappa_i = |\eta_{\mathbf{g}^i}|^2 \prod_m v_m^{-2 (g^i_m)^2}$, and $i$ enumerates different quasiparticles participating in tunneling. I remind the reader that $Q_i$ are the electric charges of the quasiparticles and $\delta$ is their common scaling dimension. The formulas \eqref{G_i}, \eqref{F_TT}, \eqref{F_0T} are correct for $\delta < 1/2$, for $\delta \geq 1/2$ they should be modified. However, typically the quasiparticles contributing to the tunneling processes are predicted to have $\delta < 1/2$.

If one applies the formulas above to analyze experimental data, one often finds a significant disagreement between the theory and experiment already for the tunneling rate $r$ (see, e.g., \cite{TunnellingRate0, TunnellingRate1, TunnellingRate2} and references therein). This is believed to be the result of non-universal physical processes in the system which can lead to (a) renormalization of the scaling dimension $\delta$ \cite{Rosenow2002, Yang2003, Papa2004, Braggio2012} and/or (b) tunneling amplitudes $\eta_{\mathbf{g}^i}$ depending on various external parameters such as the injected currents $I_s$, $I_n$, system temperature $T_0$ \cite{ShtaSniChe_2014}. Both effects are likely to be relevant in realistic situations. The tunneling amplitudes should be exponentially sensitive to the distance between the edges in the QPC since they describe tunneling of quasiparticles under a barrier. The distance between the electrostatically confined edges is in turn sensitive to the edges' electrostatic potentials, which change in the course of a tunneling experiment. The scaling dimension renormalization is also likely to be relevant in experiments. For example, the mechanism of renormalization due to 1/$f$ noise, proposed in Ref.~\cite{Braggio2012}, is extremely robust: even vanishingly small interaction of the FQHE edge with the 1/$f$ noise can produce a finite renormalization of the scaling dimension. Moreover, this mechanism (unlike the ones of Refs.~\cite{Rosenow2002, Yang2003, Papa2004}) is equally applicable for any type of the FQHE edge: with or without counterflowing modes.

However, in this work I assume that no scaling dimension renormalization happens, but the tunneling amplitudes can depend on the system parameters.\footnote{I must acknowledge here, that an unknown dependence of the tunneling amplitudes leads to a huge ambiguity: for example, one can fit any dependence of $r$ on $I_s$. The question regarding possible dependences of the tunneling amplitudes deserves a study. For example, in the case of $\nu = 2/3$ considered in Ref.~\cite{ShtaSniChe_2014} the dependence turns out not to be generic. The tunneling amplitudes for the quasiparticles there depend significantly on $I_s$ and $I_n$, but in the same way for different quasiparticles, i.e., the ratios $\kappa_i/\kappa_j$ are constant. An explanation for such a restriction is unknown to me.
\\
In contrast to this, the approach with scaling dimensions being renormalized, uses a few unknown fitting parameters, but no unknown functions. In some cases the latter approach allows one to describe the data for both tunneling rate and noise very well using a finite number of fitting parameters \cite{Ferraro2008, Ferraro, Carrega2011}. Whether it is possible to describe the case of $\nu = 2/3$ considered in Ref.~\cite{ShtaSniChe_2014} within this approach is a matter of future investigation.}

For such a case, it has been argued in \cite{ShtaSniChe_2014, Snizhko2015} that instead of considering $r$ and $\tilde{S}(0)$ separately, it is advantageous to consider their ratio (noise to tunneling rate ratio, NtTRR)
\eq{\label{noise_tun}
X(I_s) = \frac{\tilde{S}(0)}{r} = e I_s \frac{\sum_{i} \kappa_i F_i}{\sum_{i} \kappa_i G_i}.
}

In the large-$I_s$ limit one obtains
\nmq{
\label{Noise_Tun_asymp_charge}
X(I_s)\left|_{|j_s| \gg \lambda(I_n) \geq 1}\right. = e I_s \frac{\sum_{i} \kappa_i F_i}{\sum_{i} \kappa_i G_i} =\\
 = e |I_s| \frac{\sum_i \kappa_i Q_i^{4\delta+1}}{\sum_i \kappa_i Q_i^{4\delta}} + O\left(\lambda I_0, I_0\right),
}
or equivalently
\begin{eqnarray}
\label{Noise_effective_charge}
\tilde{S}(0)\left|_{|j_s| \gg \lambda(I_n) \geq 1}\right. &=& Q^* e r |I_s| = Q^* e |I_T|,\\
\label{Effective_charge_pert_def}
Q^* &=& \frac{\sum_i \kappa_i Q_i^{4\delta+1}}{\sum_i \kappa_i Q_i^{4\delta}}.
\end{eqnarray}
Therefore, in the regime of weak quasiparticle tunneling, the large-$I_s$ asymptote of the ratio of the measured excess noise and the tunneling current is equal to some average of the quasiparticle charges $Q^*$ (the coefficient $Q^* e$ is often called the Fano factor). This well-known result is correct not just for the model I consider here, but is quite robust against non-universal processes that may influence the physics at the QPC \cite{Rosenow2002}.

The average (or effective) charge $Q^*$ may be not a constant but a function of $I_s$ as the tunneling amplitudes $\eta_{\mathbf{g}^i}$ contained in $\kappa_i$ may depend on $I_s$ strongly. However, in the cases I consider in this paper this does not happen. If all the quasiparticles participating in tunneling processes have the same charge $Q_i = Q$ (as it is for the model of $\nu = 2/5$ I consider), then $Q^* = Q$ independently of the tunneling amplitudes' dependence on the current. In the case of $\nu = 2/3$ not all the $Q_i$ are equal. However, when the ratios $\kappa_i/\kappa_j$ are constant (as it has been shown \cite{ShtaSniChe_2014} for the data I consider below), then again $Q^*$ does not depend on $I_s$.

A more accurate large-$I_s$ asymptotic expression for the NtTRR, obtained in \cite{Snizhko2015},
\nmq{\label{Noise_Tun_asymp_full}
X(I_s)\left|_{|j_s| \gg \lambda(I_n) \geq 1}\right. =\\
 = e |I_s| \frac{\sum_i \kappa_i Q_i^{4\delta+1}}{\sum_i \kappa_i Q_i^{4\delta}} + e I_0 \frac{2-8\delta}{\pi} + O\left(\frac{\lambda^2 I_0^2}{I_s}, \frac{I_0^2}{I_s}\right),
}
may be useful in some cases. Its subleading term contains information about the scaling dimension of the tunneling quasiparticles.

To conclude the section, the main statements I would like the reader to take from it are as follows. The perturbative treatment of tunneling processes allows one to obtain results for experimental observables in the limit of weak quasiparticle tunneling. From large-$I_s$ asymptote of the ratio of the excess noise and the tunneling rate one can obtain some average of the tunneling quasiparticles' charges $Q^*$, which is often called the effective charge.

\subsection{\label{Sect_TheorDescr_Exact}Exact solutions}

The cases for which the model of tunneling experiments in QHE can be solved exactly are scarce: only two cases are known two me. One is the case of $\nu = 1$ integer QHE (IQHE), the other is the Laughlin sequence of states $\nu = 1/(2 k + 1)$, $k \in \mathbb{N}$. I briefly discuss these two cases below.

In the case of $\nu = 1$ IQHE the simplest edge theory is a theory of free chiral electrons. It can be rewritten into the model of a single free chiral boson of the type described in section~\ref{Sect_TunnExp_Model} through the standard bozonization technique \cite{Giamarchi2003}. The edge contains a charged mode and no neutral modes, so the injection of $I_n$ does not influence the observable quantities. The quasiparticle with the smallest scaling dimension (i.e. the one giving dominant contribution to the tunneling processes) in this case is the electron. Therefore, the problem of tunneling is a problem of free electrons that can be scattered back by a $\delta$-function barrier. Such a model can be solved exactly \cite{Lesovik1989, MartinLandauer, Buttiker1992}. The results are as follows. The tunneling rate $r = \mathrm{const}$, i.e., $r$ does not depend on $I_s$. The excess noise has the form
\eq{\label{noise_free_fermions}
\tilde{S}(0) = r(1-r) \left(e I_s \coth{\left(\frac{\pi j_s}{2}\right)} - \frac{2}{\pi}e I_0\right),
}
$j_s$ and $I_0$ are defined in Eq.~\eqref{j_s}. In the limit $r \rightarrow 0$ this expression can be reproduced by treating tunneling perturbatively as in section~\ref{Sect_TheorDescr_Pert}, with the electron as the tunneling quasiparticle (the electric charge $Q = 1$, the scaling dimension $\delta = 1/2$).

The simplest models for the FQHE edge at $\nu = 1/(2 k + 1)$, $k \in \mathbb{N}$ also contain a single charged mode. However, solving such models exactly requires the use of the Bethe ansatz technique. The details for the solution together with answers can be found in Refs.~\cite{Saleur, Saleur2, Saleur_LaughlinNoise_detailed}. Analytic answers for this case are only available for zero-temperature ($T_0 = 0$). Therefore, the use of this exact solution is not easy and requires a certain skill level in using Bethe ansatz.

\subsection{\label{Sect_TheorDescr_Phenom}Phenomenological approach}

A third, phenomenological, approach to treating the experiment model is often used in experimental papers \cite{HeiblumForFerraro, HeiblumExp, Chung2003, FracChargeObs_Heiblum, Dolev2010, Dolev2008, Griffiths2000, TunnellingRate2, FracChargeObs_Glattli}. Essentially it does not use any solution of the model but generalizes the answer of Eq.~\eqref{noise_free_fermions} for $\nu = 1$:
\eq{\label{noise_phenom}
\tilde{S}(0) = r(1-r) \left(e^* I_s \coth{\left(\frac{e^*}{e}\frac{\pi j_s}{2}\right)} - \frac{2}{\pi}e I_0\right),
}
where $j_s$ and $I_0$ are defined in Eq.~\eqref{j_s}, and $e^*$ is a phenomenological parameter.

The advantages of this approach are the simplicity of Eq.~\eqref{noise_phenom} and the correct leading asymptotic behavior:
\eq{\tilde{S}(0)\left|_{I_s = 0}\right. = 0,}
and for $r \ll 1$
\eq{
\tilde{S}(0)\left|_{|j_s| \gg 1}\right. = e^* r |I_s| = e^* |I_T|.
}
The last equation is in accord with Eq.~\eqref{Noise_effective_charge} for $e^* = e Q^*$.

Formula \eqref{noise_phenom} was compared against the exact solution for the simplest model of $\nu = 1/3$ FQHE in Ref.~\cite{SaleurGlattliCothangentNoiseValidity}, where a good agreement was found.

However, there are several disadvantages to using formula \eqref{noise_phenom}. As it has been pointed out in Ref.~\cite{Ferraro}, the value of $e^*$ extracted from real experimental data with the help of formula \eqref{noise_phenom} depends strongly on the range of $I_s$ considered. Second, the formula cannot be derived from a model for the FQHE. Indeed, the result \eqref{noise_free_fermions} is derived for the model of non-interacting electrons. If one replaces the charge of particles $e$ by $e^*$, then the edge conductance would be $(e^*)^2/h \neq \nu e^2/h$. E.g., for the Laughlin series of states $e^* = \nu e$, so that $(e^*)^2/h = \nu^2 e^2/h$. Third, the formula does not catch correctly the subleading terms in the large-$I_s$ asymptotic behavior of NtTRR \eqref{Noise_Tun_asymp_full} that carry information about the tunneling quasiparticles' scaling dimension. Finally, it does not have a natural way to include the influence of $I_n$, since such an effect is impossible in the simplest model of $\nu = 1$ IQHE.

From the above one can see that formula \eqref{noise_phenom} is a good interpolation formula, but not more. In particular, one should be careful when trying to apply it to the experiments in which injection of current $I_n$ plays an important role.

\section{\label{Sect_Nu_23}$\nu = 2/3$: neutral mode heating and behavior of the effective charge}

The $\nu = 2/3$ edge has been predicted to support a single charged mode and a single counterpropagating neutral mode \cite{KaneFisherPolchinski, KaneFisher}. The experiment of the type described in section~\ref{Sect_TunnExpScheme} that was reported in Ref.~\cite{HeiblumExp} was able to confirm qualitatively the existence of a counterflowing neutral mode. I and my co-authors analysed the experiment data quantitatively \cite{ShtaSniChe_2014} and found a good quantitative agreement between the data and the theory described in section~\ref{Sect_TheorDescr_Pert}. However, Ref.~\cite{HeiblumExp} reported a dependence of the effective charge $e^*$ on the injected current $I_n$ (see Fig.~3b of Ref.~\cite{HeiblumExp}), while in our theory the effective charge does not have such a dependence. In this section I explain the origin of this apparent discrepancy.

The $\nu = 2/3$ edge can be described with a model that contains a single charged mode and a single neutral mode that propagates opposite to the charged one. Then, according to what has been discussed in section~\ref{Sect_TunnExp_Model}, the upper edge gets heated upon injection of current $I_n$, so that the upper edge temperature near the QPC is equal to $\lambda(I_n)T_0$, while the lower edge temperature is always equal to $T_0$. There are three quasiparticles that give contribution to the tunneling processes, their charges are $Q_1 = Q_2 = 1/3$ and $Q_3 = 2/3$, their scaling dimension is $\delta = 1/3$. Using the formulas of section~\ref{Sect_TheorDescr_Pert} one can compare the theory with the experimental data for the tunneling rate $r \approx 0.2$ presented in Fig.~3a of Ref.~\cite{HeiblumExp}. In Ref.~\cite{ShtaSniChe_2014}, as a result of such comparison, it was found that the experimental data can be described well with $\lambda(I_n) = 1 + C |I_n|^a$, where $C = 5.05(13) nA^{-a}$ and $a = 0.54(5)$, and parameter $\kappa = \kappa_3/(\kappa_1 + \kappa_2)$ being independent of $I_s$ and $I_n$ and equal to $\kappa = 0.39$. The temperature $T_0 = 10$~mK is taken to be the same for all values of $I_n$. From the formulas of section~\ref{Sect_TheorDescr_Pert} one can see that the effective charge $Q^*$ is then equal to $Q^* \approx 0.5$ independently of $I_n$.

In Ref.~\cite{HeiblumExp} the experimental data is analysed with the help of Eq.~\eqref{noise_phenom} using $e^*$ and $T_0$ as fitting parameters. The interpretation connected to this is that for $I_n = 0$ the temperature $T_0$ is equal to the environment temperature, but injection of $I_n \neq 0$ heats up the whole system (or at least both edges of the QPC) leading to a different value of $T_0$ in Eq.~\eqref{noise_phenom}. The resulting agreement with the experimental data is also good, however, the behavior of the effective charge $e^*$ is different. One would expect $e^* = Q^*e/(1-r) \approx 0.63 e$ independently of $I_n$, but Fig.~3b of Ref.~\cite{HeiblumExp} says that the effective charge $e^*$ varies from $e^* \approx 2e/3$ to $e^* \approx 0.4 e$ as a function of $I_n$.

In order to explain this I fit the perturbative formula for measured current noise\footnote{A careful reader has noticed that the Nyquist noise $2 \nu e^2 k_B T_0/h$ here is two times greater than the Nyquist noise subtracted in Eq.~\eqref{excess_noise_def}. The Nyquist noise subtracted in Eq.~\eqref{excess_noise_def} is the Nyquist noise of a single edge to the left of \textit{Voltage~probe} contact in Fig.~\ref{fig_exp_scheme}. However, in a real experiment there is a similar set of edge transport channels to the right of \textit{Voltage~probe}. While these are not influenced by the injection of $I_s$ and $I_n$, they still contribute to the experimentally measured noise through the Nyquist noise. That is the origin of the Nyquist noise "doubling" here.}
\eq{\label{FullNoise_pert}
S_{\mathrm{pert}}(0) = r X(I_s) + 2 \frac{\nu e^2}{h} k_B T_0,
}
where $X(I_s)$ is defined in Eq.~\eqref{noise_tun}, with the phenomenological formula
\nmq{\label{FullNoise_phenom}
S_{\mathrm{phen}}(0) = r(1-r) \left(e^* I_s \coth{\left(\frac{e^*}{e}\frac{\pi j_s}{2}\right)} - \frac{2}{\pi}e I_0\right) \\+ 2 \frac{\nu e^2}{h} k_B T_0.
}
In other words, I repeat the analysis done in Ref.~\cite{HeiblumExp}, using the theory of Ref.~\cite{ShtaSniChe_2014} instead of experimental data.

\begin{figure}[tbp]
  \center{\includegraphics[width=1\linewidth]{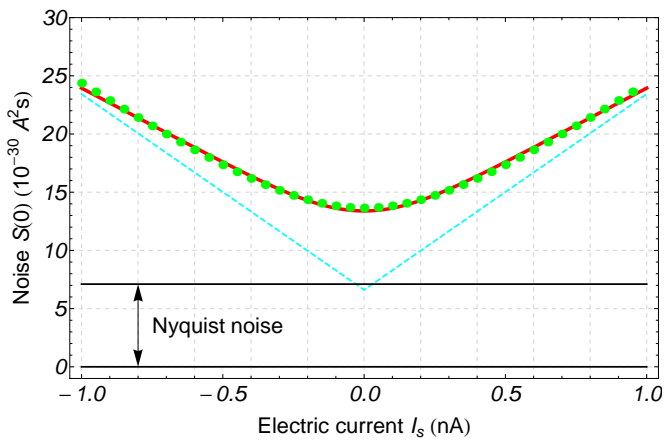}}
  \caption{\textbf{$\nu = 2/3$ measured current noise: a fit of the perturbative theory by the phenomenological formula}. (Color online). The green points are generated with the help of perturbative theory for $T_0 = 10$~mK and $\lambda(I_n) = 8$ ($I_n \approx 1.8$~nA). The red curve is a fit of these points by phenomenological formula \eqref{FullNoise_phenom}. The dashed cyan curve is the large-$I_s$ asymptote of the perturbative theory.}
  \label{fig_effective_charge_23_fit}
\end{figure}

A typical resulting fit is shown in Fig.~\ref{fig_effective_charge_23_fit}. The range of $I_s$ values corresponds to the experimental data range in Ref.~\cite{HeiblumExp}. As one can see, the fit of perturbative formula \eqref{FullNoise_pert} by phenomenological formula \eqref{FullNoise_phenom} is very good. However, the perturbative theory only starts leveling off to its large-$I_s$ asymptote \eqref{Noise_Tun_asymp_full} at $|I_s| \approx 1$~nA. Therefore, one can hardly expect the fitted $e^*$ to correspond to the proper effective charge. Indeed, the fitted value of $e^{*}_{\mathrm{fitted}} \approx 0.48 e$ differs from the expected $e^* = Q^*e/(1-r) \approx 0.63 e$. The system temperature $T_0^{\mathrm{fitted}} \approx 19$~mK given by the fit is also different from both the lower edge temperature $T_0 = 10$~mK and the upper edge temperature near the QPC $\lambda(I_n) T_0 = 80$~mK.

\begin{figure}[tbp]
  \center{\includegraphics[width=1\linewidth]{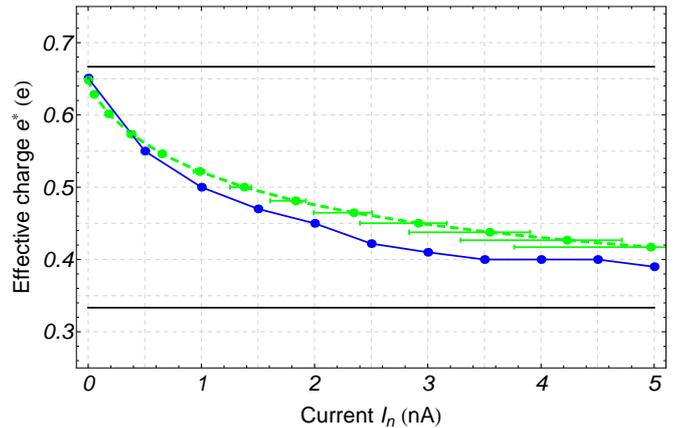}}
  \caption{\textbf{$\nu = 2/3$ measured current noise: dependence of the effective charge $e^*$ on current $I_n$}. (Color online). The blue points and the blue solid line show the data of Fig.~3b of Ref.~\cite{HeiblumExp}. The green points and the green dashed line show the dependence of $e^*_{\mathrm{fitted}}$ on $I_n$ obtained by fitting the perturbative theory with phenomenological formula \eqref{FullNoise_phenom}. The error bars are due to uncertainty in the function $\lambda(I_n)$. The black horizontal lines correspond to $e^* = 2e/3$ and $e^* = e/3$.}
  \label{fig_effective_charge_23}
\end{figure}

I repeat the fitting procedure for different values of $\lambda(I_n)$. The resulting dependence of the fitted effective charge is shown in Fig.~\ref{fig_effective_charge_23}. One can see that the dependence of the fitted effective charge $e^*_{\mathrm{fitted}}$ closely follows the data reported in Ref.~\cite{HeiblumExp}.

At the same time the Fano factor \eqref{Effective_charge_pert_def} stays constant since $\kappa = \kappa_3/(\kappa_1 + \kappa_2)$ is independent of $I_n$. This suggests that the true origin of the reported effective charge dependence on $I_n$ is the use of phenomenological formula \eqref{FullNoise_phenom} for the analysis of experimental data, and is not related to the non-universal behavior of the tunneling amplitudes, nor to complications in the $\nu = 2/3$ edge theory (for example, like the ones proposed in Refs.~\cite{Ferraro, MeirGefen23EdgeReconstruction}).\footnote{Though, the more elaborate models proposed in Refs.~\cite{Ferraro, MeirGefen23EdgeReconstruction} may be necessary to explain other observed effects. Consult the works themselves for more details.}

\section{\label{Sect_Nu_25}$\nu = 2/5$: system temperature dependence of the effective charge}

One of the other cases when an unexpected dependence of the effective charge on external parameters was reported concerns $\nu = 2/5$ \cite{Chung2003}. Namely, the effective charge was reported to depend on the system temperature $T_0$. In this section I apply the same methodology as in the previous section to this case. I find that the effective charge dependence on temperature cannot be explained the same way.

The simplest model for $\nu = 2/5$ edge contains two edge channels: the charged one and the neutral one. However, unlike in the case of $\nu = 2/3$, the neutral mode here propagates in the same direction as the charged mode. Therefore, the injection of $I_n$ should not have any effect on the measured current noise whatsoever. This was confirmed in Ref.~\cite{HeiblumExp}. Therefore, I expect that both edges have the same temperature $T_0$. There are two quasiparticles which contribute most to the tunneling processes, their electric charges are $Q_1 = Q_2 = 2/5$, their scaling dimension is $\delta = 1/5$. Therefore, the parameters $\kappa_i$ drop out of the perturbative expression for the NtTRR.

Figures~2b and 2c of Ref.~\cite{Chung2003} give some data regarding the measured current noise behavior for $r \approx 0.02$. The data of Fig.~2c of Ref.~\cite{Chung2003} presents data on the behaviour of the effective charge $e^*$ as a function of temperature $T_0$ obtained with the help of phenomenological formula~\eqref{FullNoise_phenom}.

\begin{figure}[tbp]
  \center{\includegraphics[width=1\linewidth]{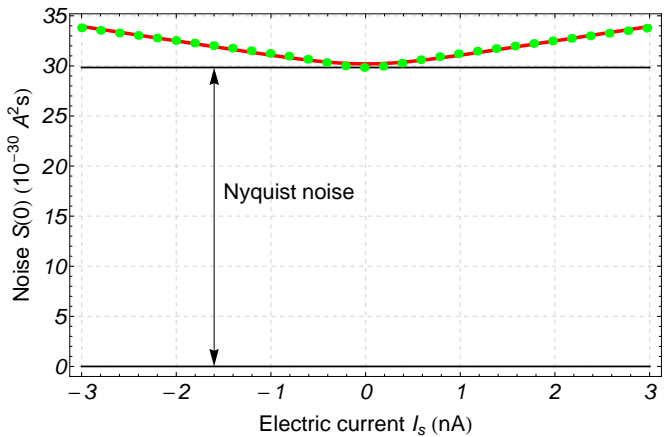}}
  \caption{\textbf{$\nu = 2/5$ measured current noise: a fit of the perturbative theory by the phenomenological formula}. (Color online). The green points are generated with the help of perturbative theory for $T_0 = 70$~mK. The red curve is a fit of these points by phenomenological formula \eqref{FullNoise_phenom}.}
  \label{fig_effective_charge_25_fit}
\end{figure}

\begin{figure}[tbp]
  \center{\includegraphics[width=1\linewidth]{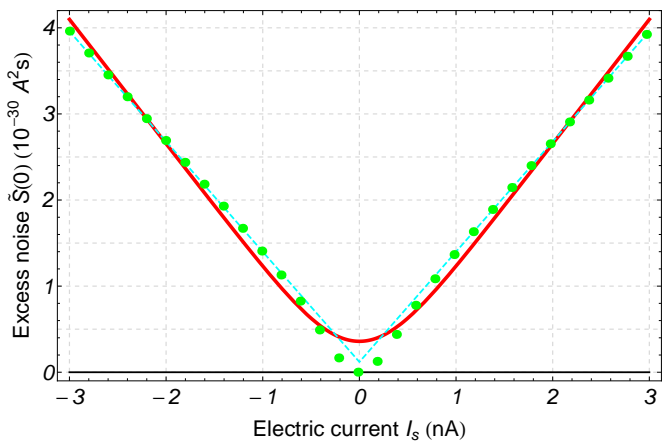}}
  \caption{\textbf{$\nu = 2/5$ measured current excess noise: a fit of the perturbative theory by the phenomenological formula}. (Color online). The green points are generated with the help of perturbative theory for $T_0 = 70$~mK. The red curve is the fitted curve from Fig.~\ref{fig_effective_charge_25_fit} less the Nyquist noise. The dashed cyan curve is the large-$I_s$ asymptote of the perturbative theory.}
  \label{fig_effective_charge_25_fit_excess}
\end{figure}

I repeat the analysis of the previous section for this case, fitting perturbative formula~\eqref{FullNoise_pert} (using the parameters corresponding to the experimental ones) with phenomenological formula~\eqref{FullNoise_phenom}. A typical resulting fit is shown in Fig.~\ref{fig_effective_charge_25_fit}. The same data with the Nyquist noise subtracted are shown in Fig.~\ref{fig_effective_charge_25_fit_excess}. As one can see, the fit is very good. However, the fitted values of the effective charge and the system temperature are slightly overestimated: $e^{*}_{\mathrm{fitted}} \approx 0.46 e$ and $T_0^{\mathrm{fitted}} \approx 71$~mK.

\begin{figure}[tbp]
  \center{\includegraphics[width=1\linewidth]{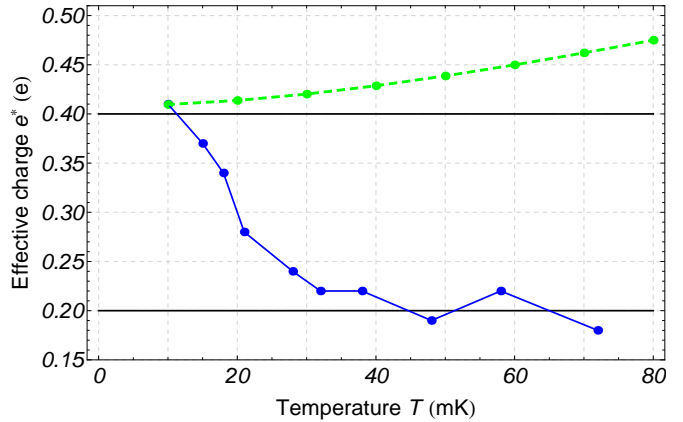}}
  \caption{\textbf{$\nu = 2/5$ measured current noise: dependence of the effective charge $e^*$ on the system temperature~$T_0$}. (Color online). The blue points and the blue solid line show the data of Fig.~2c of Ref.~\cite{Chung2003}. The green points and the green dashed line show the dependence of $e^*_{\mathrm{fitted}}$ on $T_0$ obtained by fitting the perturbative theory with phenomenological formula \eqref{FullNoise_phenom}. The black horizontal lines correspond to $e^* = 2e/5$ and $e^* = e/5$.}
  \label{fig_effective_charge_25}
\end{figure}

Repeating the fitting procedure for different values of~$T_0$, I obtain the dependence of the effective charge on temperature shown in Fig.~\ref{fig_effective_charge_25}. At the lowest considered temperature $T_0 = 10$~mK the approach can explain the oberved effective charge slightly higher than the expected $e^* = 2e/5$. However, at higher temperatures the effective charge reported in Ref.~\cite{Chung2003} drops down, while the fitted charge $e^*_{\mathrm{fitted}}$ grows.

Therefore, in this case the effective charge dependence on an external parameter cannot be explained as a peculiarity of the phenomenological formula used for the data analysis. In other words, the data of Ref.~\cite{Chung2003} regarding $\nu = 2/5$ does not agree with the perturbative theory for the simplest $\nu = 2/5$ edge model.

Finding models that can describe the data goes beyond the present article. However, I would like to mention several possibilities.

The authors of Refs.~\cite{Ferraro2008, Ferraro} showed that the data can be explained with the help of a more complicated model that (a) introduces energy cutoffs to the edge modes, (b) takes into account tunneling of the quasiparticles having the next smallest scaling dimension, (c) introduces renormalization of the scaling dimension due to non-universal processes (and assumes that tunneling amplitudes do not depend on $I_s$ and $T_0$).

While the model of Refs.~\cite{Ferraro2008, Ferraro} allows one to achieve a good quantitative agreement, it incorporates several aspects not usually considered. Therefore, it would be interesting to check whether it is possible to describe the data without some of the complications. For example, one could investigate the influence of the quasiparticles having the next smallest scaling dimension without introducing scaling dimension renormalization and/or edge mode cutoffs.

There may be other important effects. One can consider a model, in which the injection of current $I_s$ excites the copropagating neutral mode in $\nu = 2/5$ just as the injection of current $I_n$ excites the counterpropagating neutral mode in $\nu = 2/3$.

Finally, one can investigate the influence of bulk dynamics on the experimental observables. Indeed, the bias voltage corresponding to injection of the experimentally used current $I_s = 3$~nA onto the $\nu = 2/5$ edge is of the order of $200$~$\mu$V, which corresponds to energies (in the units of temperature) about~$2$~K. With the typical bulk gap in the FQHE systems on the order of (and typically less than)~$1$~K one can expect that bulk dynamics is involved at such voltages.

\section{\label{Sect_Discuss_Concl}CONCLUSION}

In this paper I compared two approaches to analyzing tunneling current noise experiments in the FQHE: the approach based on the perturbative treatment of tunneling processes in the model describing such experiments in the FQHE and the approach that uses a phenomenological generalization of the theory which describes such experiments in $\nu = 1$~IQHE.

The analysis of section~\ref{Sect_Nu_23} shows that using the phenomenological formula can lead to misinterpretation of the experimental data, like the false dependence of the effective charge on current $I_n$. However, this does not always happen, as shows the case of section~\ref{Sect_Nu_25}.

While one should be cautious when interpreting the parameters, such as the effective charge $e^*$ or system temperature $T_0$, obtained with the help of the phenomenological theory, the formula itself shows a remarkable ability to fit the proper theory well, as can be seen from sections~\ref{Sect_Nu_23}, \ref{Sect_Nu_25} and as was previously shown in Ref.~\cite{SaleurGlattliCothangentNoiseValidity}. Therefore, one can use the phenomenological formula as an efficient way to encode a large set of experimental points into two numbers $e^*$ and $T_0$. However, one can then miss subtle effects related to the scaling dimension of the quasiparticles participating in tunneling.

Finally, I would like to emphasize that the phenomenological approach is used in most papers that analyze experimental data of the tunneling current noise experiments in the FQHE, including Refs.~\cite{FracChargeObs_Heiblum, FracChargeObs_Glattli, Griffiths2000, Chung2003, Dolev2008, HeiblumForFerraro, Dolev2010, HeiblumExp}. Exceptions like Refs.~\cite{Ferraro2008, Ferraro, Carrega2011, MeirGefen23EdgeReconstruction, ShtaSniChe_2014} are rare. Therefore, reanalyzing the available experimental data with the help of proper theory can be highly beneficial. One reason is that false effects, such as in the case considered in section~\ref{Sect_Nu_23}, are possible. Another reason is that missed effects are also possible.

\section*{Acknowledgements}

I would like to thank Alessandro Braggio, Jianhui Wang, Vadim Cheianov, and Oles Shtanko for useful discussions. The research leading to these results has received funding from the European Research Council under the European Union's Seventh Framework Programme (FP7/2007-2013) / ERC grant agreement No 279738 - NEDFOQ.

\bibliography{library}% Produces the bibliography via BibTeX.

\begin{thebibliography}{37}%
\makeatletter
\providecommand \@ifxundefined [1]{%
 \@ifx{#1\undefined}
}%
\providecommand \@ifnum [1]{%
 \ifnum #1\expandafter \@firstoftwo
 \else \expandafter \@secondoftwo
 \fi
}%
\providecommand \@ifx [1]{%
 \ifx #1\expandafter \@firstoftwo
 \else \expandafter \@secondoftwo
 \fi
}%
\providecommand \natexlab [1]{#1}%
\providecommand \enquote  [1]{``#1''}%
\providecommand \bibnamefont  [1]{#1}%
\providecommand \bibfnamefont [1]{#1}%
\providecommand \citenamefont [1]{#1}%
\providecommand \href@noop [0]{\@secondoftwo}%
\providecommand \href [0]{\begingroup \@sanitize@url \@href}%
\providecommand \@href[1]{\@@startlink{#1}\@@href}%
\providecommand \@@href[1]{\endgroup#1\@@endlink}%
\providecommand \@sanitize@url [0]{\catcode `\\12\catcode `\$12\catcode
  `\&12\catcode `\#12\catcode `\^12\catcode `\_12\catcode `\%12\relax}%
\providecommand \@@startlink[1]{}%
\providecommand \@@endlink[0]{}%
\providecommand \url  [0]{\begingroup\@sanitize@url \@url }%
\providecommand \@url [1]{\endgroup\@href {#1}{\urlprefix }}%
\providecommand \urlprefix  [0]{URL }%
\providecommand \Eprint [0]{\href }%
\providecommand \doibase [0]{http://dx.doi.org/}%
\providecommand \selectlanguage [0]{\@gobble}%
\providecommand \bibinfo  [0]{\@secondoftwo}%
\providecommand \bibfield  [0]{\@secondoftwo}%
\providecommand \translation [1]{[#1]}%
\providecommand \BibitemOpen [0]{}%
\providecommand \bibitemStop [0]{}%
\providecommand \bibitemNoStop [0]{.\EOS\space}%
\providecommand \EOS [0]{\spacefactor3000\relax}%
\providecommand \BibitemShut  [1]{\csname bibitem#1\endcsname}%
\let\auto@bib@innerbib\@empty
%</preamble>
\bibitem [{\citenamefont {Nayak}\ \emph {et~al.}(2008)\citenamefont {Nayak},
  \citenamefont {Simon}, \citenamefont {Stern}, \citenamefont {Freedman},\ and\
  \citenamefont {{Das Sarma}}}]{TopolQuantCompFQHE}%
  \BibitemOpen
  \bibfield  {author} {\bibinfo {author} {\bibfnamefont {C.}~\bibnamefont
  {Nayak}}, \bibinfo {author} {\bibfnamefont {S.~H.}\ \bibnamefont {Simon}},
  \bibinfo {author} {\bibfnamefont {A.}~\bibnamefont {Stern}}, \bibinfo
  {author} {\bibfnamefont {M.}~\bibnamefont {Freedman}}, \ and\ \bibinfo
  {author} {\bibfnamefont {S.}~\bibnamefont {{Das Sarma}}},\ }\href {\doibase
  10.1103/RevModPhys.80.1083} {\bibfield  {journal} {\bibinfo  {journal}
  {Reviews of Modern Physics}\ }\textbf {\bibinfo {volume} {80}},\ \bibinfo
  {pages} {1083} (\bibinfo {year} {2008})}\BibitemShut {NoStop}%
\bibitem [{\citenamefont {Kane}\ and\ \citenamefont {Fisher}(1994)}]{Kane1994}%
  \BibitemOpen
  \bibfield  {author} {\bibinfo {author} {\bibfnamefont {C.~L.}\ \bibnamefont
  {Kane}}\ and\ \bibinfo {author} {\bibfnamefont {M.~P.~A.}\ \bibnamefont
  {Fisher}},\ }\href {http://link.aps.org/doi/10.1103/PhysRevLett.72.724}
  {\bibfield  {journal} {\bibinfo  {journal} {Physical Review Letters}\
  }\textbf {\bibinfo {volume} {72}},\ \bibinfo {pages} {724} (\bibinfo {year}
  {1994})}\BibitemShut {NoStop}%
\bibitem [{\citenamefont {De-Picciotto}\ \emph {et~al.}(1997)\citenamefont
  {De-Picciotto}, \citenamefont {Reznikov}, \citenamefont {Heiblum},
  \citenamefont {Umansky}, \citenamefont {Bunin},\ and\ \citenamefont
  {Mahalu}}]{FracChargeObs_Heiblum}%
  \BibitemOpen
  \bibfield  {author} {\bibinfo {author} {\bibfnamefont {R.}~\bibnamefont
  {De-Picciotto}}, \bibinfo {author} {\bibfnamefont {M.}~\bibnamefont
  {Reznikov}}, \bibinfo {author} {\bibfnamefont {M.}~\bibnamefont {Heiblum}},
  \bibinfo {author} {\bibfnamefont {V.}~\bibnamefont {Umansky}}, \bibinfo
  {author} {\bibfnamefont {G.}~\bibnamefont {Bunin}}, \ and\ \bibinfo {author}
  {\bibfnamefont {D.}~\bibnamefont {Mahalu}},\ }\href
  {http://dx.doi.org/10.1038/38241} {\bibfield  {journal} {\bibinfo  {journal}
  {Nature}\ }\textbf {\bibinfo {volume} {389}},\ \bibinfo {pages} {162}
  (\bibinfo {year} {1997})}\BibitemShut {NoStop}%
\bibitem [{\citenamefont {Saminadayar}\ \emph {et~al.}(1997)\citenamefont
  {Saminadayar}, \citenamefont {Glattli}, \citenamefont {Jin},\ and\
  \citenamefont {Etienne}}]{FracChargeObs_Glattli}%
  \BibitemOpen
  \bibfield  {author} {\bibinfo {author} {\bibfnamefont {L.}~\bibnamefont
  {Saminadayar}}, \bibinfo {author} {\bibfnamefont {D.~C.}\ \bibnamefont
  {Glattli}}, \bibinfo {author} {\bibfnamefont {Y.}~\bibnamefont {Jin}}, \ and\
  \bibinfo {author} {\bibfnamefont {B.}~\bibnamefont {Etienne}},\ }\href
  {http://dx.doi.org/10.1103/PhysRevLett.79.2526} {\bibfield  {journal}
  {\bibinfo  {journal} {Physical Review Letters}\ }\textbf {\bibinfo {volume}
  {79}},\ \bibinfo {pages} {2526} (\bibinfo {year} {1997})}\BibitemShut
  {NoStop}%
\bibitem [{\citenamefont {Griffiths}\ \emph {et~al.}(2000)\citenamefont
  {Griffiths}, \citenamefont {Comforti}, \citenamefont {Heiblum}, \citenamefont
  {Stern},\ and\ \citenamefont {Umansky}}]{Griffiths2000}%
  \BibitemOpen
  \bibfield  {author} {\bibinfo {author} {\bibfnamefont {T.~G.}\ \bibnamefont
  {Griffiths}}, \bibinfo {author} {\bibfnamefont {E.}~\bibnamefont {Comforti}},
  \bibinfo {author} {\bibfnamefont {M.}~\bibnamefont {Heiblum}}, \bibinfo
  {author} {\bibfnamefont {A.}~\bibnamefont {Stern}}, \ and\ \bibinfo {author}
  {\bibfnamefont {V.}~\bibnamefont {Umansky}},\ }\href {\doibase
  10.1103/PhysRevLett.85.3918} {\bibfield  {journal} {\bibinfo  {journal}
  {Physical Review Letters}\ }\textbf {\bibinfo {volume} {85}},\ \bibinfo
  {pages} {3918} (\bibinfo {year} {2000})}\BibitemShut {NoStop}%
\bibitem [{\citenamefont {Chung}\ \emph {et~al.}(2003)\citenamefont {Chung},
  \citenamefont {Heiblum},\ and\ \citenamefont {Umansky}}]{Chung2003}%
  \BibitemOpen
  \bibfield  {author} {\bibinfo {author} {\bibfnamefont {Y.~C.}\ \bibnamefont
  {Chung}}, \bibinfo {author} {\bibfnamefont {M.}~\bibnamefont {Heiblum}}, \
  and\ \bibinfo {author} {\bibfnamefont {V.}~\bibnamefont {Umansky}},\ }\href
  {http://dx.doi.org/10.1103/PhysRevLett.91.216804} {\bibfield  {journal}
  {\bibinfo  {journal} {Physical Review Letters}\ }\textbf {\bibinfo {volume}
  {91}},\ \bibinfo {pages} {216804} (\bibinfo {year} {2003})}\BibitemShut
  {NoStop}%
\bibitem [{\citenamefont {Dolev}\ \emph {et~al.}(2008)\citenamefont {Dolev},
  \citenamefont {Heiblum}, \citenamefont {Umansky}, \citenamefont {Stern},\
  and\ \citenamefont {Mahalu}}]{Dolev2008}%
  \BibitemOpen
  \bibfield  {author} {\bibinfo {author} {\bibfnamefont {M.}~\bibnamefont
  {Dolev}}, \bibinfo {author} {\bibfnamefont {M.}~\bibnamefont {Heiblum}},
  \bibinfo {author} {\bibfnamefont {V.}~\bibnamefont {Umansky}}, \bibinfo
  {author} {\bibfnamefont {A.}~\bibnamefont {Stern}}, \ and\ \bibinfo {author}
  {\bibfnamefont {D.}~\bibnamefont {Mahalu}},\ }\href
  {http://dx.doi.org/10.1038/nature06855} {\bibfield  {journal} {\bibinfo
  {journal} {Nature}\ }\textbf {\bibinfo {volume} {452}},\ \bibinfo {pages}
  {829} (\bibinfo {year} {2008})}\BibitemShut {NoStop}%
\bibitem [{\citenamefont {Bid}\ \emph {et~al.}(2009)\citenamefont {Bid},
  \citenamefont {Ofek}, \citenamefont {Heiblum}, \citenamefont {Umansky},\ and\
  \citenamefont {Mahalu}}]{HeiblumForFerraro}%
  \BibitemOpen
  \bibfield  {author} {\bibinfo {author} {\bibfnamefont {A.}~\bibnamefont
  {Bid}}, \bibinfo {author} {\bibfnamefont {N.}~\bibnamefont {Ofek}}, \bibinfo
  {author} {\bibfnamefont {M.}~\bibnamefont {Heiblum}}, \bibinfo {author}
  {\bibfnamefont {V.}~\bibnamefont {Umansky}}, \ and\ \bibinfo {author}
  {\bibfnamefont {D.}~\bibnamefont {Mahalu}},\ }\href
  {http://link.aps.org/doi/10.1103/PhysRevLett.103.236802} {\bibfield
  {journal} {\bibinfo  {journal} {Physical Review Letters}\ }\textbf {\bibinfo
  {volume} {103}},\ \bibinfo {pages} {236802} (\bibinfo {year}
  {2009})}\BibitemShut {NoStop}%
\bibitem [{\citenamefont {Dolev}\ \emph {et~al.}(2010)\citenamefont {Dolev},
  \citenamefont {Gross}, \citenamefont {Chung}, \citenamefont {Heiblum},
  \citenamefont {Umansky},\ and\ \citenamefont {Mahalu}}]{Dolev2010}%
  \BibitemOpen
  \bibfield  {author} {\bibinfo {author} {\bibfnamefont {M.}~\bibnamefont
  {Dolev}}, \bibinfo {author} {\bibfnamefont {Y.}~\bibnamefont {Gross}},
  \bibinfo {author} {\bibfnamefont {Y.~C.}\ \bibnamefont {Chung}}, \bibinfo
  {author} {\bibfnamefont {M.}~\bibnamefont {Heiblum}}, \bibinfo {author}
  {\bibfnamefont {V.}~\bibnamefont {Umansky}}, \ and\ \bibinfo {author}
  {\bibfnamefont {D.}~\bibnamefont {Mahalu}},\ }\href
  {http://dx.doi.org/10.1103/PhysRevB.81.161303} {\bibfield  {journal}
  {\bibinfo  {journal} {Physical Review B}\ }\textbf {\bibinfo {volume} {81}},\
  \bibinfo {pages} {161303} (\bibinfo {year} {2010})}\BibitemShut {NoStop}%
\bibitem [{\citenamefont {Bid}\ \emph {et~al.}(2010)\citenamefont {Bid},
  \citenamefont {Ofek}, \citenamefont {Inoue}, \citenamefont {Heiblum},
  \citenamefont {Kane}, \citenamefont {Umansky},\ and\ \citenamefont
  {Mahalu}}]{HeiblumExp}%
  \BibitemOpen
  \bibfield  {author} {\bibinfo {author} {\bibfnamefont {A.}~\bibnamefont
  {Bid}}, \bibinfo {author} {\bibfnamefont {N.}~\bibnamefont {Ofek}}, \bibinfo
  {author} {\bibfnamefont {H.}~\bibnamefont {Inoue}}, \bibinfo {author}
  {\bibfnamefont {M.}~\bibnamefont {Heiblum}}, \bibinfo {author} {\bibfnamefont
  {C.~L.}\ \bibnamefont {Kane}}, \bibinfo {author} {\bibfnamefont
  {V.}~\bibnamefont {Umansky}}, \ and\ \bibinfo {author} {\bibfnamefont
  {D.}~\bibnamefont {Mahalu}},\ }\href {http://dx.doi.org/10.1038/nature09277}
  {\bibfield  {journal} {\bibinfo  {journal} {Nature}\ }\textbf {\bibinfo
  {volume} {466}},\ \bibinfo {pages} {585} (\bibinfo {year}
  {2010})}\BibitemShut {NoStop}%
\bibitem [{\citenamefont {Ferraro}\ \emph {et~al.}(2008)\citenamefont
  {Ferraro}, \citenamefont {Braggio}, \citenamefont {Merlo}, \citenamefont
  {Magnoli},\ and\ \citenamefont {Sassetti}}]{Ferraro2008}%
  \BibitemOpen
  \bibfield  {author} {\bibinfo {author} {\bibfnamefont {D.}~\bibnamefont
  {Ferraro}}, \bibinfo {author} {\bibfnamefont {A.}~\bibnamefont {Braggio}},
  \bibinfo {author} {\bibfnamefont {M.}~\bibnamefont {Merlo}}, \bibinfo
  {author} {\bibfnamefont {N.}~\bibnamefont {Magnoli}}, \ and\ \bibinfo
  {author} {\bibfnamefont {M.}~\bibnamefont {Sassetti}},\ }\href
  {http://dx.doi.org/10.1103/PhysRevLett.101.166805} {\bibfield  {journal}
  {\bibinfo  {journal} {Physical Review Letters}\ }\textbf {\bibinfo {volume}
  {101}},\ \bibinfo {pages} {166805} (\bibinfo {year} {2008})}\BibitemShut
  {NoStop}%
\bibitem [{\citenamefont {Ferraro}\ \emph {et~al.}(2010)\citenamefont
  {Ferraro}, \citenamefont {Braggio}, \citenamefont {Magnoli},\ and\
  \citenamefont {Sassetti}}]{Ferraro}%
  \BibitemOpen
  \bibfield  {author} {\bibinfo {author} {\bibfnamefont {D.}~\bibnamefont
  {Ferraro}}, \bibinfo {author} {\bibfnamefont {A.}~\bibnamefont {Braggio}},
  \bibinfo {author} {\bibfnamefont {N.}~\bibnamefont {Magnoli}}, \ and\
  \bibinfo {author} {\bibfnamefont {M.}~\bibnamefont {Sassetti}},\ }\href
  {http://link.aps.org/doi/10.1103/PhysRevB.82.085323} {\bibfield  {journal}
  {\bibinfo  {journal} {Physical Review B}\ }\textbf {\bibinfo {volume} {82}},\
  \bibinfo {pages} {085323} (\bibinfo {year} {2010})}\BibitemShut {NoStop}%
\bibitem [{\citenamefont {Carrega}\ \emph {et~al.}(2011)\citenamefont
  {Carrega}, \citenamefont {Ferraro}, \citenamefont {Braggio}, \citenamefont
  {Magnoli},\ and\ \citenamefont {Sassetti}}]{Carrega2011}%
  \BibitemOpen
  \bibfield  {author} {\bibinfo {author} {\bibfnamefont {M.}~\bibnamefont
  {Carrega}}, \bibinfo {author} {\bibfnamefont {D.}~\bibnamefont {Ferraro}},
  \bibinfo {author} {\bibfnamefont {A.}~\bibnamefont {Braggio}}, \bibinfo
  {author} {\bibfnamefont {N.}~\bibnamefont {Magnoli}}, \ and\ \bibinfo
  {author} {\bibfnamefont {M.}~\bibnamefont {Sassetti}},\ }\href
  {http://dx.doi.org/10.1103/PhysRevLett.107.146404} {\bibfield  {journal}
  {\bibinfo  {journal} {Physical Review Letters}\ }\textbf {\bibinfo {volume}
  {107}},\ \bibinfo {pages} {146404} (\bibinfo {year} {2011})}\BibitemShut
  {NoStop}%
\bibitem [{\citenamefont {Wang}\ \emph {et~al.}(2013)\citenamefont {Wang},
  \citenamefont {Meir},\ and\ \citenamefont
  {Gefen}}]{MeirGefen23EdgeReconstruction}%
  \BibitemOpen
  \bibfield  {author} {\bibinfo {author} {\bibfnamefont {J.}~\bibnamefont
  {Wang}}, \bibinfo {author} {\bibfnamefont {Y.}~\bibnamefont {Meir}}, \ and\
  \bibinfo {author} {\bibfnamefont {Y.}~\bibnamefont {Gefen}},\ }\href
  {http://dx.doi.org/10.1103/PhysRevLett.111.246803} {\bibfield  {journal}
  {\bibinfo  {journal} {Physical Review Letters}\ }\textbf {\bibinfo {volume}
  {111}},\ \bibinfo {pages} {246803} (\bibinfo {year} {2013})}\BibitemShut
  {NoStop}%
\bibitem [{\citenamefont {Shtanko}\ \emph {et~al.}(2014)\citenamefont
  {Shtanko}, \citenamefont {Snizhko},\ and\ \citenamefont
  {Cheianov}}]{ShtaSniChe_2014}%
  \BibitemOpen
  \bibfield  {author} {\bibinfo {author} {\bibfnamefont {O.}~\bibnamefont
  {Shtanko}}, \bibinfo {author} {\bibfnamefont {K.}~\bibnamefont {Snizhko}}, \
  and\ \bibinfo {author} {\bibfnamefont {V.}~\bibnamefont {Cheianov}},\ }\href
  {http://link.aps.org/doi/10.1103/PhysRevB.89.125104} {\bibfield  {journal}
  {\bibinfo  {journal} {Physical Review B}\ }\textbf {\bibinfo {volume} {89}},\
  \bibinfo {pages} {125104} (\bibinfo {year} {2014})}\BibitemShut {NoStop}%
\bibitem [{\citenamefont {Heiblum}(2006)}]{TunnellingRate2}%
  \BibitemOpen
  \bibfield  {author} {\bibinfo {author} {\bibfnamefont {M.}~\bibnamefont
  {Heiblum}},\ }\href {http://dx.doi.org/10.1002/pssb.200642237} {\bibfield
  {journal} {\bibinfo  {journal} {physica status solidi (b)}\ }\textbf
  {\bibinfo {volume} {243}},\ \bibinfo {pages} {3604} (\bibinfo {year}
  {2006})}\BibitemShut {NoStop}%
\bibitem [{\citenamefont {Trauzettel}\ \emph {et~al.}(2004)\citenamefont
  {Trauzettel}, \citenamefont {Roche}, \citenamefont {Glattli},\ and\
  \citenamefont {Saleur}}]{SaleurGlattliCothangentNoiseValidity}%
  \BibitemOpen
  \bibfield  {author} {\bibinfo {author} {\bibfnamefont {B.}~\bibnamefont
  {Trauzettel}}, \bibinfo {author} {\bibfnamefont {P.}~\bibnamefont {Roche}},
  \bibinfo {author} {\bibfnamefont {D.~C.}\ \bibnamefont {Glattli}}, \ and\
  \bibinfo {author} {\bibfnamefont {H.}~\bibnamefont {Saleur}},\ }\href
  {http://link.aps.org/doi/10.1103/PhysRevB.70.233301} {\bibfield  {journal}
  {\bibinfo  {journal} {Physical Review B}\ }\textbf {\bibinfo {volume} {70}},\
  \bibinfo {pages} {233301} (\bibinfo {year} {2004})}\BibitemShut {NoStop}%
\bibitem [{\citenamefont {Kane}\ \emph {et~al.}(1994)\citenamefont {Kane},
  \citenamefont {Fisher},\ and\ \citenamefont
  {Polchinski}}]{KaneFisherPolchinski}%
  \BibitemOpen
  \bibfield  {author} {\bibinfo {author} {\bibfnamefont {C.~L.}\ \bibnamefont
  {Kane}}, \bibinfo {author} {\bibfnamefont {M.~P.~A.}\ \bibnamefont {Fisher}},
  \ and\ \bibinfo {author} {\bibfnamefont {J.}~\bibnamefont {Polchinski}},\
  }\href {http://link.aps.org/doi/10.1103/PhysRevLett.72.4129} {\bibfield
  {journal} {\bibinfo  {journal} {Physical Review Letters}\ }\textbf {\bibinfo
  {volume} {72}},\ \bibinfo {pages} {4129} (\bibinfo {year}
  {1994})}\BibitemShut {NoStop}%
\bibitem [{\citenamefont {Kane}\ and\ \citenamefont
  {Fisher}(1995)}]{KaneFisher}%
  \BibitemOpen
  \bibfield  {author} {\bibinfo {author} {\bibfnamefont {C.~L.}\ \bibnamefont
  {Kane}}\ and\ \bibinfo {author} {\bibfnamefont {M.~P.~A.}\ \bibnamefont
  {Fisher}},\ }\href {http://link.aps.org/doi/10.1103/PhysRevB.51.13449}
  {\bibfield  {journal} {\bibinfo  {journal} {Physical Review B}\ }\textbf
  {\bibinfo {volume} {51}},\ \bibinfo {pages} {13449} (\bibinfo {year}
  {1995})}\BibitemShut {NoStop}%
\bibitem [{\citenamefont {Fr{\"{o}}hlich}\ \emph {et~al.}(2001)\citenamefont
  {Fr{\"{o}}hlich}, \citenamefont {Pedrini}, \citenamefont {Schweigert},\ and\
  \citenamefont {Walcher}}]{Frohlich_CosetHall}%
  \BibitemOpen
  \bibfield  {author} {\bibinfo {author} {\bibfnamefont {J.}~\bibnamefont
  {Fr{\"{o}}hlich}}, \bibinfo {author} {\bibfnamefont {B.}~\bibnamefont
  {Pedrini}}, \bibinfo {author} {\bibfnamefont {C.}~\bibnamefont {Schweigert}},
  \ and\ \bibinfo {author} {\bibfnamefont {J.}~\bibnamefont {Walcher}},\ }\href
  {http://link.springer.com/article/10.1023/A:1010389232079} {\bibfield
  {journal} {\bibinfo  {journal} {Journal of Statistical Physics}\ }\textbf
  {\bibinfo {volume} {103}},\ \bibinfo {pages} {527} (\bibinfo {year}
  {2001})}\BibitemShut {NoStop}%
\bibitem [{\citenamefont {Wen}(1990)}]{WenCLL}%
  \BibitemOpen
  \bibfield  {author} {\bibinfo {author} {\bibfnamefont {X.~G.}\ \bibnamefont
  {Wen}},\ }\href {http://link.aps.org/doi/10.1103/PhysRevB.41.12838}
  {\bibfield  {journal} {\bibinfo  {journal} {Physical Review B}\ }\textbf
  {\bibinfo {volume} {41}},\ \bibinfo {pages} {12838} (\bibinfo {year}
  {1990})}\BibitemShut {NoStop}%
\bibitem [{\citenamefont {Fr{\"{o}}hlich}\ and\ \citenamefont
  {Kerler}(1991)}]{FrohlichAnomaly}%
  \BibitemOpen
  \bibfield  {author} {\bibinfo {author} {\bibfnamefont {J.}~\bibnamefont
  {Fr{\"{o}}hlich}}\ and\ \bibinfo {author} {\bibfnamefont {T.}~\bibnamefont
  {Kerler}},\ }\href {http://dx.doi.org/10.1016/0550-3213(91)90360-A}
  {\bibfield  {journal} {\bibinfo  {journal} {Nuclear Physics B}\ }\textbf
  {\bibinfo {volume} {354}},\ \bibinfo {pages} {369} (\bibinfo {year}
  {1991})}\BibitemShut {NoStop}%
\bibitem [{\citenamefont {Fendley}\ \emph
  {et~al.}(1995{\natexlab{a}})\citenamefont {Fendley}, \citenamefont {Ludwig},\
  and\ \citenamefont {Saleur}}]{Saleur}%
  \BibitemOpen
  \bibfield  {author} {\bibinfo {author} {\bibfnamefont {P.}~\bibnamefont
  {Fendley}}, \bibinfo {author} {\bibfnamefont {A.~W.~W.}\ \bibnamefont
  {Ludwig}}, \ and\ \bibinfo {author} {\bibfnamefont {H.}~\bibnamefont
  {Saleur}},\ }\href {http://link.aps.org/doi/10.1103/PhysRevB.52.8934}
  {\bibfield  {journal} {\bibinfo  {journal} {Physical Review B}\ }\textbf
  {\bibinfo {volume} {52}},\ \bibinfo {pages} {8934} (\bibinfo {year}
  {1995}{\natexlab{a}})}\BibitemShut {NoStop}%
\bibitem [{\citenamefont {Fendley}\ \emph
  {et~al.}(1995{\natexlab{b}})\citenamefont {Fendley}, \citenamefont {Ludwig},\
  and\ \citenamefont {Saleur}}]{Saleur2}%
  \BibitemOpen
  \bibfield  {author} {\bibinfo {author} {\bibfnamefont {P.}~\bibnamefont
  {Fendley}}, \bibinfo {author} {\bibfnamefont {A.~W.~W.}\ \bibnamefont
  {Ludwig}}, \ and\ \bibinfo {author} {\bibfnamefont {H.}~\bibnamefont
  {Saleur}},\ }\href {http://dx.doi.org/10.1103/PhysRevLett.75.2196} {\bibfield
   {journal} {\bibinfo  {journal} {Physical Review Letters}\ }\textbf {\bibinfo
  {volume} {75}},\ \bibinfo {pages} {2196} (\bibinfo {year}
  {1995}{\natexlab{b}})}\BibitemShut {NoStop}%
\bibitem [{\citenamefont {Huntington}\ and\ \citenamefont
  {Cheianov}()}]{HuntingtonCheianov_TunnAmplitudeMonteCarlo}%
  \BibitemOpen
  \bibfield  {author} {\bibinfo {author} {\bibfnamefont {S.}~\bibnamefont
  {Huntington}}\ and\ \bibinfo {author} {\bibfnamefont {V.}~\bibnamefont
  {Cheianov}},\ }\href@noop {} {\ }\Eprint {http://arxiv.org/abs/1410.6638}
  {arXiv:1410.6638} \BibitemShut {NoStop}%
\bibitem [{\citenamefont {Lesovik}(1989)}]{Lesovik1989}%
  \BibitemOpen
  \bibfield  {author} {\bibinfo {author} {\bibfnamefont {G.~B.}\ \bibnamefont
  {Lesovik}},\ }\href
  {http://www.jetpletters.ac.ru/ps/1120/article{\_}16970.shtml} {\bibfield
  {journal} {\bibinfo  {journal} {JETP Letters}\ }\textbf {\bibinfo {volume}
  {49}},\ \bibinfo {pages} {592} (\bibinfo {year} {1989})}\BibitemShut
  {NoStop}%
\bibitem [{\citenamefont {Martin}\ and\ \citenamefont
  {Landauer}(1992)}]{MartinLandauer}%
  \BibitemOpen
  \bibfield  {author} {\bibinfo {author} {\bibfnamefont {T.}~\bibnamefont
  {Martin}}\ and\ \bibinfo {author} {\bibfnamefont {R.}~\bibnamefont
  {Landauer}},\ }\href {http://link.aps.org/doi/10.1103/PhysRevB.45.1742}
  {\bibfield  {journal} {\bibinfo  {journal} {Physical Review B}\ }\textbf
  {\bibinfo {volume} {45}},\ \bibinfo {pages} {1742} (\bibinfo {year}
  {1992})}\BibitemShut {NoStop}%
\bibitem [{\citenamefont {Snizhko}\ and\ \citenamefont
  {Cheianov}(2015)}]{Snizhko2015}%
  \BibitemOpen
  \bibfield  {author} {\bibinfo {author} {\bibfnamefont {K.}~\bibnamefont
  {Snizhko}}\ and\ \bibinfo {author} {\bibfnamefont {V.}~\bibnamefont
  {Cheianov}},\ }\href {http://link.aps.org/doi/10.1103/PhysRevB.91.195151}
  {\bibfield  {journal} {\bibinfo  {journal} {Physical Review B}\ }\textbf
  {\bibinfo {volume} {91}},\ \bibinfo {pages} {195151} (\bibinfo {year}
  {2015})}\BibitemShut {NoStop}%
\bibitem [{\citenamefont {Chang}(2003)}]{TunnellingRate0}%
  \BibitemOpen
  \bibfield  {author} {\bibinfo {author} {\bibfnamefont {A.~M.}\ \bibnamefont
  {Chang}},\ }\href {http://link.aps.org/doi/10.1103/RevModPhys.75.1449}
  {\bibfield  {journal} {\bibinfo  {journal} {Reviews of Modern Physics}\
  }\textbf {\bibinfo {volume} {75}},\ \bibinfo {pages} {1449} (\bibinfo {year}
  {2003})}\BibitemShut {NoStop}%
\bibitem [{\citenamefont {Grayson}\ \emph {et~al.}(1998)\citenamefont
  {Grayson}, \citenamefont {Tsui}, \citenamefont {Pfeiffer}, \citenamefont
  {West},\ and\ \citenamefont {Chang}}]{TunnellingRate1}%
  \BibitemOpen
  \bibfield  {author} {\bibinfo {author} {\bibfnamefont {M.}~\bibnamefont
  {Grayson}}, \bibinfo {author} {\bibfnamefont {D.~C.}\ \bibnamefont {Tsui}},
  \bibinfo {author} {\bibfnamefont {L.~N.}\ \bibnamefont {Pfeiffer}}, \bibinfo
  {author} {\bibfnamefont {K.~W.}\ \bibnamefont {West}}, \ and\ \bibinfo
  {author} {\bibfnamefont {A.~M.}\ \bibnamefont {Chang}},\ }\href
  {http://link.aps.org/doi/10.1103/PhysRevLett.80.1062} {\bibfield  {journal}
  {\bibinfo  {journal} {Physical Review Letters}\ }\textbf {\bibinfo {volume}
  {80}},\ \bibinfo {pages} {1062} (\bibinfo {year} {1998})}\BibitemShut
  {NoStop}%
\bibitem [{\citenamefont {Rosenow}\ and\ \citenamefont
  {Halperin}(2002)}]{Rosenow2002}%
  \BibitemOpen
  \bibfield  {author} {\bibinfo {author} {\bibfnamefont {B.}~\bibnamefont
  {Rosenow}}\ and\ \bibinfo {author} {\bibfnamefont {B.~I.}\ \bibnamefont
  {Halperin}},\ }\href {http://dx.doi.org/10.1103/PhysRevLett.88.096404}
  {\bibfield  {journal} {\bibinfo  {journal} {Physical Review Letters}\
  }\textbf {\bibinfo {volume} {88}},\ \bibinfo {pages} {96404} (\bibinfo {year}
  {2002})}\BibitemShut {NoStop}%
\bibitem [{\citenamefont {Yang}(2003)}]{Yang2003}%
  \BibitemOpen
  \bibfield  {author} {\bibinfo {author} {\bibfnamefont {K.}~\bibnamefont
  {Yang}},\ }\href {http://link.aps.org/doi/10.1103/PhysRevLett.91.036802}
  {\bibfield  {journal} {\bibinfo  {journal} {Physical Review Letters}\
  }\textbf {\bibinfo {volume} {91}},\ \bibinfo {pages} {36802} (\bibinfo {year}
  {2003})}\BibitemShut {NoStop}%
\bibitem [{\citenamefont {Papa}\ and\ \citenamefont
  {MacDonald}(2004)}]{Papa2004}%
  \BibitemOpen
  \bibfield  {author} {\bibinfo {author} {\bibfnamefont {E.}~\bibnamefont
  {Papa}}\ and\ \bibinfo {author} {\bibfnamefont {A.~H.}\ \bibnamefont
  {MacDonald}},\ }\href {http://dx.doi.org/10.1103/PhysRevLett.93.126801}
  {\bibfield  {journal} {\bibinfo  {journal} {Physical Review Letters}\
  }\textbf {\bibinfo {volume} {93}},\ \bibinfo {pages} {126801} (\bibinfo
  {year} {2004})}\BibitemShut {NoStop}%
\bibitem [{\citenamefont {Braggio}\ \emph {et~al.}(2012)\citenamefont
  {Braggio}, \citenamefont {Ferraro}, \citenamefont {Carrega}, \citenamefont
  {Magnoli},\ and\ \citenamefont {Sassetti}}]{Braggio2012}%
  \BibitemOpen
  \bibfield  {author} {\bibinfo {author} {\bibfnamefont {A.}~\bibnamefont
  {Braggio}}, \bibinfo {author} {\bibfnamefont {D.}~\bibnamefont {Ferraro}},
  \bibinfo {author} {\bibfnamefont {M.}~\bibnamefont {Carrega}}, \bibinfo
  {author} {\bibfnamefont {N.}~\bibnamefont {Magnoli}}, \ and\ \bibinfo
  {author} {\bibfnamefont {M.}~\bibnamefont {Sassetti}},\ }\href {\doibase
  10.1088/1367-2630/14/9/093032} {\bibfield  {journal} {\bibinfo  {journal}
  {New Journal of Physics}\ }\textbf {\bibinfo {volume} {14}},\ \bibinfo
  {pages} {093032} (\bibinfo {year} {2012})}\BibitemShut {NoStop}%
\bibitem [{\citenamefont {Giamarchi}(2003)}]{Giamarchi2003}%
  \BibitemOpen
  \bibfield  {author} {\bibinfo {author} {\bibfnamefont {T.}~\bibnamefont
  {Giamarchi}},\ }\href {https://books.google.com.ua/books?id=0CGVxiyUZYYC}
  {\emph {\bibinfo {title} {{Quantum Physics in One Dimension}}}},\
  International Series of Monographs on Physics\ (\bibinfo  {publisher}
  {Clarendon Press},\ \bibinfo {year} {2003})\BibitemShut {NoStop}%
\bibitem [{\citenamefont {B{\"{u}}ttiker}(1992)}]{Buttiker1992}%
  \BibitemOpen
  \bibfield  {author} {\bibinfo {author} {\bibfnamefont {M.}~\bibnamefont
  {B{\"{u}}ttiker}},\ }\href
  {http://link.aps.org/doi/10.1103/PhysRevB.46.12485} {\bibfield  {journal}
  {\bibinfo  {journal} {Physical Review B}\ }\textbf {\bibinfo {volume} {46}},\
  \bibinfo {pages} {12485} (\bibinfo {year} {1992})}\BibitemShut {NoStop}%
\bibitem [{\citenamefont {Fendley}\ and\ \citenamefont
  {Saleur}(1996)}]{Saleur_LaughlinNoise_detailed}%
  \BibitemOpen
  \bibfield  {author} {\bibinfo {author} {\bibfnamefont {P.}~\bibnamefont
  {Fendley}}\ and\ \bibinfo {author} {\bibfnamefont {H.}~\bibnamefont
  {Saleur}},\ }\href {http://dx.doi.org/10.1103/PhysRevB.54.10845} {\bibfield
  {journal} {\bibinfo  {journal} {Physical Review B}\ }\textbf {\bibinfo
  {volume} {54}},\ \bibinfo {pages} {10845} (\bibinfo {year}
  {1996})}\BibitemShut {NoStop}%
\end{thebibliography}%
\end{document}